\documentclass[acmsmall]{acmart}
\usepackage{graphicx}
\usepackage{makecell}
\usepackage{colortbl}
\usepackage{array}
\usepackage{url}
\usepackage{breakurl}
\usepackage{longtable}
\usepackage{CJKutf8}
\usepackage{gensymb}

\AtBeginDocument{%
  \providecommand\BibTeX{{%
    \normalfont B\kern-0.5em{\scshape i\kern-0.25em b}\kern-0.8em\TeX}}}

\setcopyright{acmcopyright}
\copyrightyear{2023}
\acmYear{2023}
\acmDOI{XXXXXXX.XXXXXXX}

\begin{document}


\title{Deep Learning and Medical Imaging for COVID-19 Diagnosis: A Comprehensive Survey}
\author{Song Wu}
\email{SongWu.uestc@outlook.com}
\orcid{0000-0003-0147-1907}
\author{Yazhou Ren}
\authornote{Corresponding author. E-mail: yazhou.ren@uestc.edu.cn}
\email{yazhou.ren@uestc.edu.cn}
\author{Aodi Yang}
\email{aodi.yang@outlook.com}
\author{Xinyue Chen}
\email{martinachen2580@gmail.com}
\author{Xiaorong Pu}
\email{puxiaor@uestc.edu.cn}
\affiliation{%
  \institution{School of Computer Science and Engineering, Shenzhen Institute for Advanced Study, University of Electronic Science and Technology of China}
  \country{China}
}
\author{Jing He}
\email{lotusjing@gmail.com}
\affiliation{%
  \institution{Nuffield Department of Clinical Neurosciences, University of Oxford}
  \country{UK}
}

\author{Liqiang Nie}
\email{nieliqiang@gmail.com}
\affiliation{%
  \institution{School of Computer Science and Technology, Harbin Institute of Technology (Shenzhen)}
  \country{China}
}
\author{Philip S. Yu}
\email{psyu@uic.edu}
\affiliation{%
  \institution{Department of Computer Science, University of Illinois at Chicago}
  \country{USA}
}

\renewcommand{\shortauthors}{Wu et al.}

\begin{abstract}
COVID-19 (Coronavirus disease 2019) has been quickly spreading since its outbreak, impacting financial markets and healthcare systems globally. Countries all around the world have adopted a number of extraordinary steps to restrict the spreading virus, where early COVID-19 diagnosis is essential. Medical images such as X-ray images and Computed Tomography scans are becoming one of the main diagnostic tools to combat COVID-19 with the aid of deep learning-based systems. In this survey, we investigate the main contributions of deep learning applications using medical images in fighting against COVID-19 from the aspects of image classification, lesion localization, and severity quantification, and review different deep learning architectures and some image preprocessing techniques for achieving a preciser diagnosis. We also provide a summary of the X-ray and CT image datasets used in various studies for COVID-19 detection. The key difficulties and potential applications of deep learning in fighting against COVID-19 are finally discussed. This work summarizes the latest methods of deep learning using medical images to diagnose COVID-19, highlighting the challenges and inspiring more studies to keep utilizing the advantages of deep learning to combat COVID-19.
\end{abstract}

\ccsdesc[500]{Computing methodologies~Artificial intelligence}
\ccsdesc[500]{Applied computing~Life and medical sciences}
\ccsdesc[500]{General and reference~Surveys and overviews}

\keywords{Deep learning, Medical image, COVID-19}


\maketitle

\section{Introduction}
\begin{figure}[htbp]
  \centering
  \includegraphics[width=\linewidth]{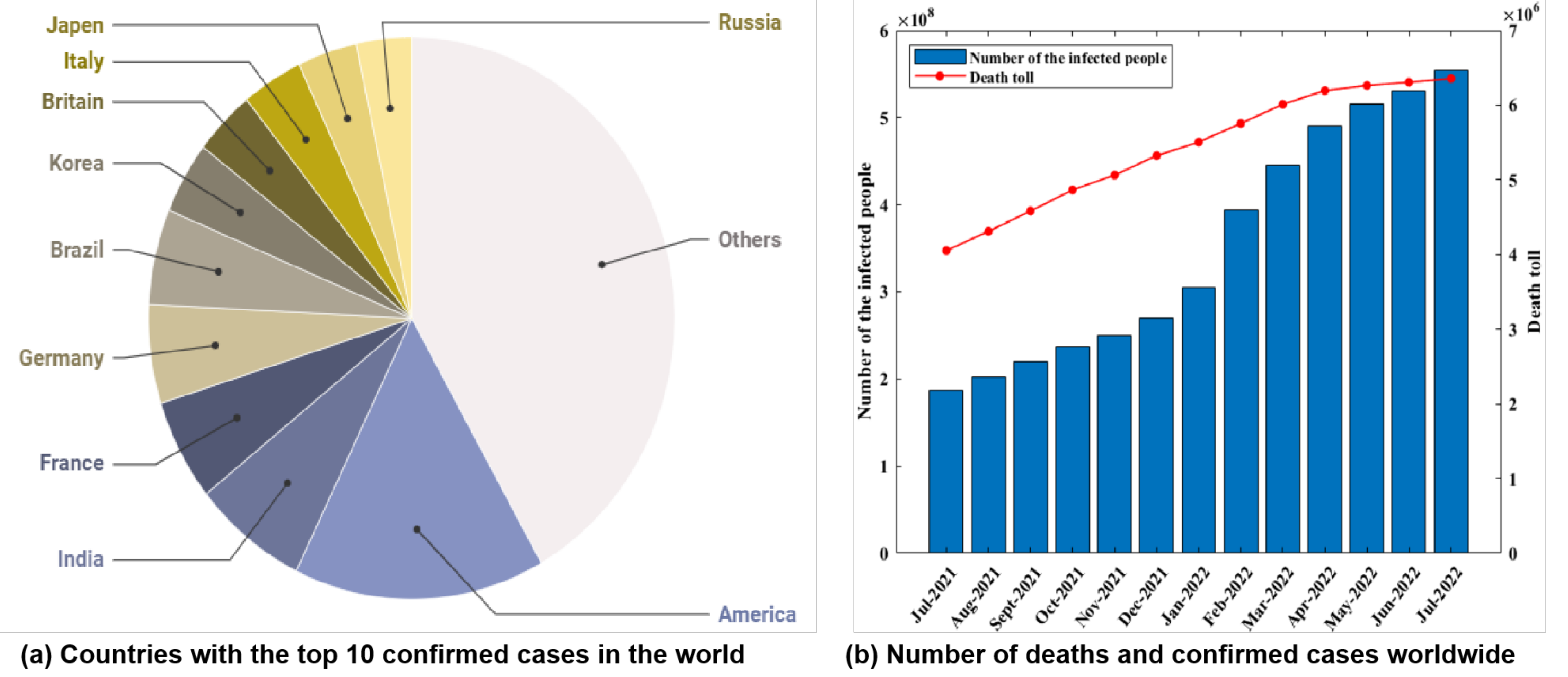}
  \caption{The number of infections and deaths in the world, and the top 10 infected countries (October 22, 2022).}
  \label{figure:COVID-19 number}
\end{figure}

\textbf{COVID-19 (Coronavirus Disease 2019)}, an acute respiratory infection caused by a novel coronavirus, was first reported in Wuhan, China, in December 2019 \cite{ref117,ref137}. It is highly contagious and the \textbf{World Health Organization (WHO)} declared that the outbreak was a \textbf{Public Health Emergency of International Concern (PHEIC)} on January 30, 2020 and one and a half month later, a pandemic. The number of confirmed and death cases has increased rapidly since its outbreak as shown in Figure \ref{figure:COVID-19 number} \cite{ref240}. As of July 31, 2022, more than 574 million confirmed cases and more than 6.3 million fatalities had been documented, severely impacting people’s daily lives. The healthcare systems of many countries are perilously close to collapse, and the world financial market is suffered a negative impact \cite{ref138,ref139,ref140,ref141}. To break the viral cycle of transmission, early screening and severity evaluation for COVID-19 patients are essential \cite{ref142,ref128,ref127}.

\textbf{RT-PCR (Reverse Transcription-Polymerase Chain Reaction)} is a technology that combines \textbf{reverse transcription (RT)} of RNA and \textbf{polymerase chain reaction (PCR)} of cDNA \cite{ref143,ref144}. Due to its high sensitivity, wide dynamic range, and accurate detection, the RT-PCR is widely used in identifying COVID-19 subjects \cite{ref235,ref236,ref121}. It, however, has a lot of limitations. First, the quality of sample collection has a significant impact on RT-PCR results, and the false negative rate is extremely high \cite{ref119,ref51,ref19,ref46}. Second, the reaction time for RT-PCR is generally lengthy, and the standards for testing certification are rather high \cite{ref145}. Third, the suspected patients are vulnerable to cross-infection during the RT-PCR test, which will exacerbate the spreading virus \cite{ref146}.

To solve the above problem, medical imaging is an available diagnostic method, which can be regarded as an important complement to RT-PCR in COVID-19 detection. It can signal lesion manifestations of different infected levels, aiding doctors to make an appropriate diagnosis\cite{ref123,ref237}. Meanwhile, many studies attempt to use deep learning technologies and medical imaging for COVID-19 early screening, which can greatly prevent the further spread of the epidemic. And the DL-based diagnostic methods have shown great potential in image classification, lesion localization and severity evaluation \cite{ref158,ref84,ref145,ref148,ref179}.

There are some surveys about using deep learning and medical imaging to fight against COVID-19 \cite{ref126,ref238,ref239,ref178}. Bhattacharya et al. \cite{ref126} summarize recent studies related to the applications of deep learning based on medical image processing and discuss their potential to combat COVID-19. 
Hryniewska et al. \cite{ref238} summarize the deep neural network methods using medical images for COVID-19 diagnosis and show how to use interpretable AI techniques in these models. Soomro et al. \cite{ref178} perform the image analysis based on CT and X-rays images from both traditional image system and AI-system directions, and summarize the various applications of AI in COVID-19 diagnosis. Liu et al. \cite{ref239} review existing deep learning models and medical image analysis methods for COVID-19 diagnosis, relating them to hot issues in deep learning research, including interpretable deep learning, and fair deep learning. While most of these surveys focus on the various applications of deep learning using medical images in fighting against COVID-19, there is a lack of discussion on how to utilize deep learning technologies to achieve a preciser diagnosis. Therefore, it is meaningful to summarize recent studies using deep learning and medical imaging to diagnose COVID-19, and explore how to utilize existing deep learning models to achieve more accurate diagnosis.

The contributions of this paper are summarized as follows:
\begin{itemize}
\item[$\bullet$]
First, we gather open source medical image datasets used in various studies, which mainly contain CT and X-ray images, to assist new studies in finding relevant and reliable medical images quickly.
\end{itemize}

\begin{itemize}
\item[$\bullet$]
Second, we systematically summarize various applications of deep learning using medical images for COVID-19 diagnosis. Different deep learning architectures and the methods used to improve the performance of deep models are reviewed.
\end{itemize}

\begin{itemize}
\item[$\bullet$]
Third, we discuss the limitations and future work of using deep learning techniques to diagnose COVID-19. And we hope that deep learning can play a crucial role in fighting against COVID-19 in the future.
\end{itemize}

\section{COVID-19 Dataset and Medical Image Manifestations}
\textbf{Deep learning (DL)} has demonstrated extraordinary capabilities in medical image processing. The performance of deep learning models, however, is significantly influenced by the quality of annotated data \cite{ref170,ref149,ref150}. A large amount of well-annotated data can improve network performance and avoid overfitting \cite{ref126,ref171}. Due to the sudden outbreak of the epidemic, the available datasets are limited and still in the development stage \cite{ref151}. Hence, it is a difficult but significant work to collect open source medical image datasets. In this section, we summarize various medical image datasets used for COVID-19 diagnosis in different papers, mainly containing CT and X-ray images. Additionally, we describe the manifestations of CT and X-ray images as well as the lesion information provided by these images.

\subsection{CT manifestations of COVID-19 and resource description}
\begin{figure}[htbp]
  \centering
  \includegraphics[width=\linewidth]{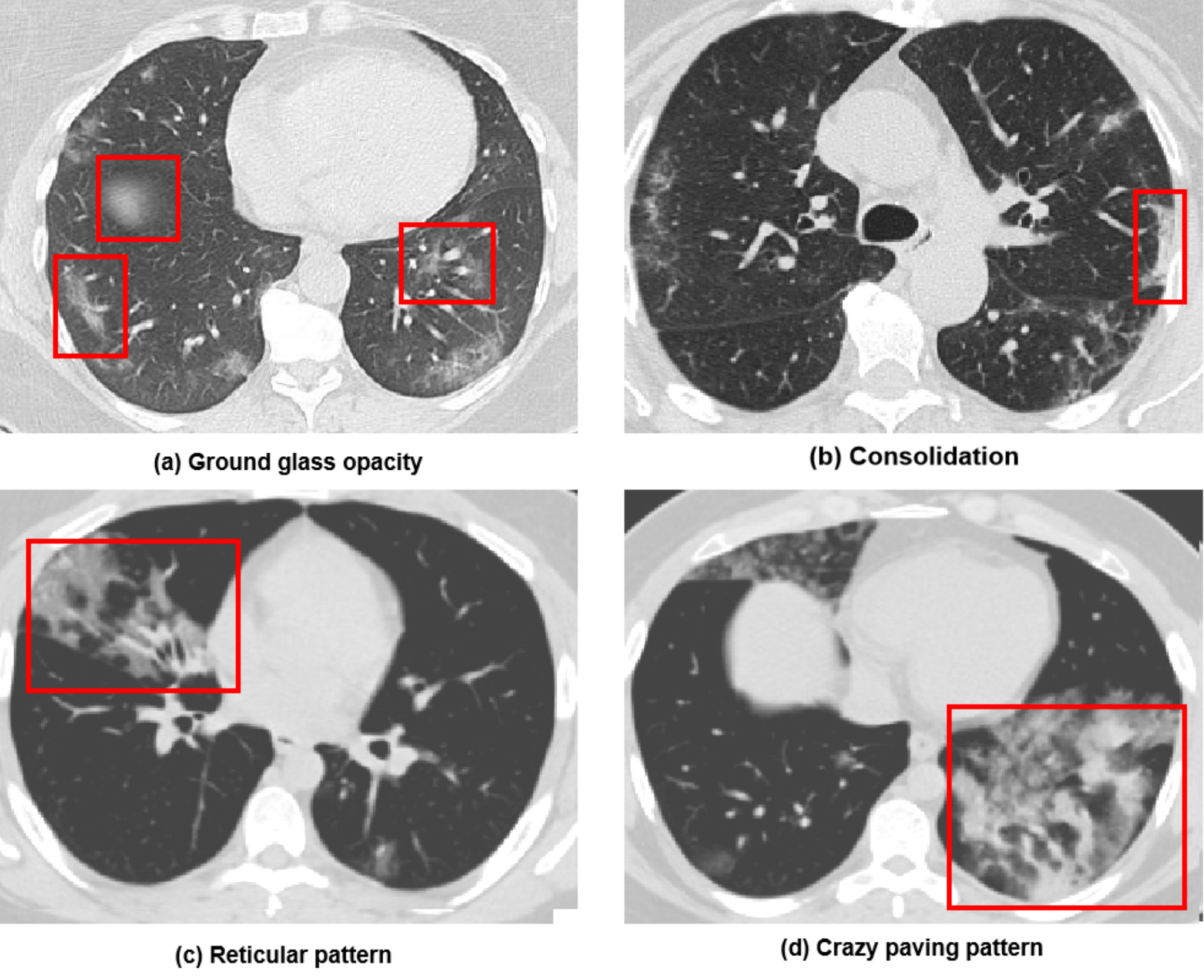}
  \caption{Typical CT image manifestations of COVID-19 patients, including (a) GGO (b) Consolidation (c) Reticular pattern (d) Crazy paving pattern.}
  \label{firgure:Lung CT images}
\end{figure}

\textbf{Computed tomography (CT)} is one of the landmark advances in medical technology, providing radiologists with a useful imaging tool. Some pathological manifestations, such as \textbf{ground glass opacity (GGO)} and \textbf{consolidation (C)}, are typically presenting in COVID-19 patients \cite{ref122,ref152}. Additionally, chest CT images may show diverse radiological characteristics or patterns as the state of COVID-19 patients evolves. In the early stage of the disease, the lesions mainly are presented as ground glass opacity (GGO) and thickened small blood vessels. Consolidation may have occurred in some cases \cite{ref153}. The patients with severe disease start to develop into a reticular pattern with interlobular septal thickening, and consolidation shadows will appear in peripheral and central regions of their lungs \cite{ref129}. When the patient's condition deteriorates further, the white lung is visible in CT images and the lesions display multiple diffuse GGO and consolidation. Typical CT manifestations of COVID-19 patients are shown in Figure \ref{firgure:Lung CT images}.

Many studies attempt to diagnose COVID-19 using CT images because of the rich pathological manifestations they provide, and deep learning models based on CT images have demonstrated outstanding performance on COVID-19 diagnosis \cite{ref131,ref154,ref155}. To explore these CT images, Nivetha et al. \cite{ref1} use a public COVID-19 CT scan dataset, which contains 349 positive COVID-19 cases and 397 negative chest CT images. Their proposed NRNN model achieves the accuracy of 0.98, 0.92, 1.00, 1.00 respectively in the experiment. Choudhary et al. \cite{ref8} use a publicly available SARS-CoV2 CT-scan dataset, which consists of 1,252 positive COVID-19 cases and 1,230 negative chest CT images. Their pruned ResNet-34 model reaches 0.9547 accuracy, 0.9216 sensitivity, 0.9567 F-score, and 0.9942 specificity. Ahrabi et al. \cite{ref6} establish a training dataset composed of 4,000 CT images of COVID-19 cases collected from more than 500 patients. Their proposed approach reaches 0.9712 accuracy, 0.9741 precision, 0.9659 recall and 0.9696 F-score. 

Additionally, some of the most popular datasets come from authoritative hospitals. Song et al. \cite{ref58} collect their dataset from three hospitals, i.e., Renmin Hospital of Wuhan University, and two affiliated hospitals of the Sun Yat-sen University in Guangzhou. Shi et al. \cite{ref131} build their dataset from three hospitals, i.e., Tongji Hospital of Huazhong University of Science and Technology, Shanghai Public Health Clinical Center of Fudan University, and China–Japan Union Hospital of Jilin University. In order to make it easier for new studies to obtain available and authoritative COVID-19 CT datasets, various datasets used in different papers are gathered and listed in Table \ref{table:dataset}. 
\begin{table}[h]
\centering
\caption{A Summary of different datasets used in the papers for COVID-19 detection.}
\resizebox{\textwidth}{!}{
\begin{tabular}{cp{4cm}p{7cm}c}
\toprule
Type & Dataset name & Dataset source & Papers     \\ 
\midrule
CT	& SARS-COV-2 Ct-Scan Dataset	& \url{https://www.kaggle.com/datasets/plameneduardo/sarscov2-ctscan-dataset}	&\cite{ref202}\\
CT	& COVID-CT	& \url{https://github.com/UCSD-AI4H/COVID-CT}	& \cite{ref222}\\
CT	& COVIDx CT	& \url{https://www.kaggle.com/datasets/hgunraj/covidxct} &	\cite{ref224}\\
CT	& Chest CT-Scan images Dataset	& \url{https://www.kaggle.com/datasets/mohamedhanyyy/chest-ctscan-images} &	\cite{ref6}\\
CT	& COVID-19 CT segmentation dataset	& \url{http://medicalsegmentation.com/COVID19/} &	\cite{ref223}\\
CT	& COVID-19	& \url{https://radiopaedia.org/articles/covid-19-3} &	\cite{ref226}\\
CT	& NSCC	& \url{https://ai.nscc-tj.cn/thai/deploy/public/pneumonia ct} & 	\cite{ref234}\\
X-ray	& COVID-19 X rays	& \url{https://www.kaggle.com/datasets/andrewmvd/convid19-x-rays} &	\cite{ref225}\\
X-ray	& COVID-chestxray-dataset	& \url{https://github.com/ieee8023/covid-chestxray-dataset} &	\cite{ref52,ref227}\\
X-ray	& Chest X-ray Images (Pneumonia)	& \url{https://www.kaggle.com/datasets/paultimothymooney/chest-xray-pneumonia} &  \cite{ref17}\\
X-ray	& COVID-19 Radiography Database	& \url{https://www.kaggle.com/datasets/tawsifurrahman/covid19-radiography-database} &	\cite{ref68}\\
X-ray	& Open Source COVID-19	& \url{https://github.com/WeileiZeng/Open-Source-COVID-19} &	\cite{ref229}\\
X-ray	& RSNA Pneumonia Detection Challenge	& \url{https://www.kaggle.com/c/rsna-pneumonia-detection-challenge/data} &	\cite{ref18}\\
X-ray	& ActualMed COVID-19 chest X-ray dataset 	& \url{https://github.com/agchung/Actualmed-COVID-chestxray-dataset} &	\cite{ref228}\\
X-ray	& COVID-19 database	& \url{https://www.sirm.org/category/senza-categoria/COVID-19/} & 	\cite{ref231}\\
X-ray	& COVID-CAPS	& \url{https://github.com/ShahinSHH/COVID-CAPS} & \cite{ref32}\\
X-ray	& NIH Chest X-rays	& \url{https://www.kaggle.com/datasets/nih-chest-xrays/data} &  \cite{ref232}\\
X-ray	& COVID-19	& \url{https://github.com/muhammedtalo/COVID-19} & \cite{ref233}\\

\bottomrule
\end{tabular}
}
\label{table:dataset}
\end{table}

\subsection{X-ray manifestations of COVID-19 and resource description}
\begin{figure}[h]
  \centering
  \includegraphics[width=\linewidth]{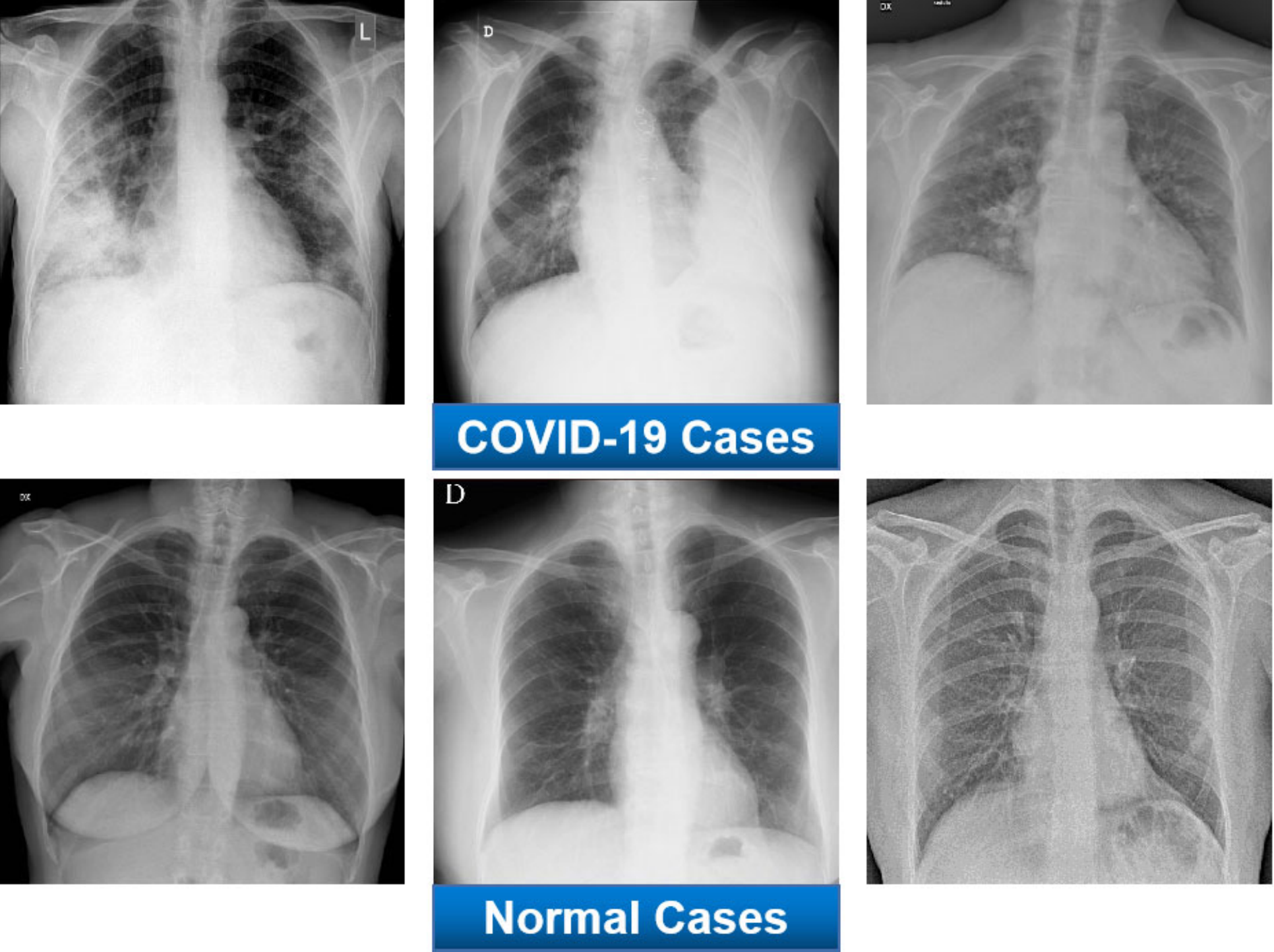}
  \caption{The representative X-ray images of COVID-19 cases and the normal cases.}
  \label{chest X-ray images}
\end{figure}

Even though CT scans provide radiologists with more information about lesions, they are more expensive and expose the patients to more radiation. The normal and COVID-19 X-ray images are shown in Figure \ref{chest X-ray images}. The \textbf{peripheral lung opacity (PLO)} and \textbf{ground glass opacity (GGO)} are two of these abnormalities \cite{ref132,ref157}. The chest X-ray images of COVID-19 patients may seem normal in the early or mild stage. About 10 to 12 days later, symptoms become more obvious. The lesions are generally distributed among the lower, peripheral, and bilateral regions \cite{ref133,ref134}. In reference \cite{ref136}, they retrospectively evaluate X-ray images of 593 patients admitted to hospital with COVID-19. The experiments demonstrate that abnormalities are mostly found in the middle and lower regions (88\% of cases), and ground glass opacity (GGO) is the most common finding in abnormal X-rays (75\% of cases). The next most common findings are peripheral lung opacity (PLO) and confluent consolidation, which generally indicate more severe symptoms.

The X-ray image datasets are more available than CT image datasets because capturing X-ray images is more convenient and less expensive than CT images. Hence, many studies attempt to use X-ray images rather than CT scans to diagnose COVID-19 \cite{ref155,ref98}. Regarding these X-ray images, Dev et al. \cite{ref18} use the COVIDx dataset, which consists of 55,328 normal cases, 8,066 bacterial pneumonia cases and 358 COVID-19 cases. Their HCN-FM and HCN-DML reach 95.33\% accuracy and 96.67\% accuracy, respectively. Annavarapu et al. \cite{ref29} use a publicly accessible COVID-19 chest X-ray dataset, which consists of 2,905 images. Their model obtains 95\% accuracy and 97\% specificity. Turkoglu \cite{ref48} collects 6,092 X-ray images from a combination of public datasets. Their proposed model reaches 99.18\% accuracy, surpassing most previous studies. 
Miao et al. \cite{ref17} use three public COVID-19 X-ray datasets and other pneumonia datasets to construct two experimental datasets. Jain et al. \cite{ref47} collect 6,432 chest X-ray images from the Kaggle repository. Various COVID-19 X-ray datasets used in different papers are listed in Table \ref{table:dataset} to facilitate new studies to quickly find available resources.

\section{Image Preprocessing Technologies}
Most COVID-19 datasets contain some blurry, repetitive, and unrelated images, which will degrade the performance of deep model \cite{ref119}. Image preprocessing is generally the first step in diagnosing COVID-19 using deep learning technologies. It eliminates irrelevant information in the image and recovers useful and real information, thereby improving the reliability and performance of diagnostic algorithms \cite{ref159,ref203}. Nivetha et al. \cite{ref1} use the median filter for the CT images of COVID-19 patients from a publicly accessible dataset. In this manner, the quality of COVID images is improved, and the noise is reduced without losing important information. Joshi et al. \cite{ref3} resize all images to 224×224 to maintain data consistency. And four types of transformation are employed to preprocess images. Ahrabi et al. \cite{ref6} use 4,000 CT images of COVID-19 cases from two publicly accessible datasets. All the images are cropped and resized to remove irrelevant regions. In this section, we will discuss extensively utilized image preprocessing technologies and their characteristics.

\begin{itemize}
\item[$\bullet$]
\textbf{Resizing} 

Resizing is an essential step for image preprocessing phase in the applications of deep learning. It uniforms the size of images from different datasets. Tao et al. \cite{ref19} resize all CT images to a uniform size (64×64) to build the COVID-19 dataset for model training. Shah et al. \cite{ref24} resize CT images from COVID-19 cases to 128×128×3 or 224×224×3 to suit the input size of different deep models. Garain et al. \cite{ref25} resize each image to a dimension of 32×32 and the images are converted to greyscale.
\end{itemize}

\begin{itemize}
\item[$\bullet$]
\textbf{Flipping and Rotating} 

Flipping and rotating are the commonly used data augmentation methods, which can guarantee that the deep learning model is trained with images at any angle and thus can make better prediction \cite{ref29}.
And the sufficient images can also help deep model avoid overfitting. Sedik et al. \cite{ref23} use a dataset consisting of 288 COVID-19 cases and 288 normal cases. The dataset is augmented by several rotations and scaling operations. Keles et al. \cite{ref36} employ three augmentation methods, including  horizontal/vertical flipping, rotation and shifting. Ahuja et al. \cite{ref45} randomly rotate the training data between range [-90\degree, 90\degree] to resolve the overfitting problem. 
\end{itemize}

\begin{itemize}
\item[$\bullet$]
\textbf{Cropping} 

Cropping refers to cropping the given image to remove irrelevant regions. It greatly reduces the interference of irrelevant information and largely increases the performance and robustness of deep learning models. Balaha et al. \cite{ref15} preprocess the image data to a suitable format for CNN to further learn the latent features. The images are cropped using a bounding rectangle to reserve the region of interest. Madaan et al. \cite{ref16} use two chest X-ray image datasets in their model. Cropping is employed to reduce the noise and all images are resized. Zhang et al. \cite{ref34} crop the rulers on the right side of the image and texts on the bottom of the image, which are irrelevant information and can impair the model performance.  
\end{itemize}

\begin{itemize}
\item[$\bullet$]
\textbf{Contrast adjusting} 

Medical images may have poor contrast during imaging process and data acquisition. It is essential to enhance image contrast to help deep models extract meaningful information from the region of interest. Dhaka et al. \cite{ref43} propose that changing the contrast of X-ray images is capable of improving the performance of deep model. Tahir et al. \cite{ref9} use contrast limited adaptive histogram equalization (CLAHE) to enhance the contrast of the original X-ray image. The produced images are more natural compared with other contrast enhancement methods. Habib et al. \cite{ref13} use histogram equalization (HE) and CLAHE to enhance the contrast.
\end{itemize}

\begin{itemize}
\item[$\bullet$]
\textbf{Denoising} 

Image noise means the unnecessary or unwanted interference information that exists in an image. Noise can seriously affect the quality of images. Hence, it is necessary to do denoising   before the downstream image processing task.
Mahendran et al. \cite{ref12} use DnCNN Algorithm to denoise CT Images before image classification. Balaha et al. \cite{ref15} convert all images to grayscale images and apply Gaussian blurring to eliminate unnecessary noise. Mishra et al. \cite{ref27} use a Gaussian filter to smooth the image and the background noise can be eliminated via a two-dimensional median filter.
\end{itemize}

Applying appropriate preprocessing methods needs to consider multiple factors such as the size of the dataset, image quality, input of the model, and the application scenarios. We investigate various papers using deep learning and medical imaging to diagnose COVID over the past three years. We summarize the representative preprocessing technologies in Table \ref{table:preprocessing technologies}.

\begin{table}[h]
\centering
\caption{A summary of various preprocessing techniques used in different papers.}
\resizebox{\textwidth}{!}{
\begin{tabular}{cp{8cm}c} 
\toprule
Preprocessing Method  & Papers\centering  & Count  \\
\midrule
Resizing & \cite{ref1,ref4,ref6,ref14,ref15,ref16,ref19,ref24,ref25,ref27,ref28,ref29,ref30,ref33,ref34,ref38,ref39,ref40,ref43,ref45,ref71,ref74,ref59,ref216,ref218,ref215,ref217,ref209} & 28   \\
Flipping and Rotating & \cite{ref3,ref16,ref23,ref27,ref29,ref32,ref34,ref36,ref41,ref44,ref45,ref214,ref218,ref215,ref209} &15 \\
Cropping & \cite{ref6,ref15,ref16,ref27,ref34,ref41,ref74,ref214,ref218,ref217,ref210} & 11\\
Contrast adjusting & \cite{ref9,ref13,ref27,ref43,ref215} & 5\\
Denoising & \cite{ref1,ref12,ref15,ref27,ref217} &5\\
\bottomrule
\end{tabular}
}
\label{table:preprocessing technologies}
\end{table}

\section{Image Segmentation}
Accurate segmentation of lung and infection in medical images of COVID-19 patients plays a crucial role in the diagnosis of COVID-19 patients \cite{ref63,ref74}. In specific, image segmentation identifies the region of interest (RoI) from given medical images and divides the medical image into several parts \cite{ref204,ref177}. The segmented region can aid doctors and deep learning models more comfortably to understand lung disease type and severity \cite{ref83,ref82,ref185}. 
Punn et al. \cite{ref75} propose a deep learning based semantic hierarchical segmenter (CHS-Net) to identify the COVID-19 infected regions from chest CT images. They apply a residual inception U-Net model with spectral spatial and depth attention network (SSD) to effectively encode and decode the semantic and varying image information, and finally segment the COVID-19 infected regions in the lungs. 
Wang et al. \cite{ref205} propose a lesion edge detection model (COVID Edge-Net), which consists of one edge detection backbone and two novel modules, i.e., the multi-scale residual dual attention (MSRDA) module and the Canny operator module. The MSRDA module can capture rich contextual associations to generate enhanced feature representations, which are then merged with Canny features learned from the Canny operator module to extract more accurate, clearer, and sharper edges. 

Chen et al. \cite{ref206} propose a segmentation network based on  unsupervised domain adaptation (UDA) to improve the segmentation performance of the infected regions in CT images. 
The generalization ability of the segmentation network is significantly enhanced by a novel domain adaption module that aligns the two domains. Their proposed method achieves high performance in lung segmentation, and it is proved that their proposed model is also suitable for large-area organs or tissues. Lashchenova et al. \cite{ref64} argue that an important factor affecting the real performance of the segmentation model is that the model segments the puncta (connected components) of lung and the lung lesions outside the real lung. To detect this, connected components are computed for every class and for each component to found whether it has interception with ground-truth lungs. 

\subsection{U-net Architecture}
\begin{figure}[htbp]
  \centering
  \includegraphics[width=0.9\linewidth]{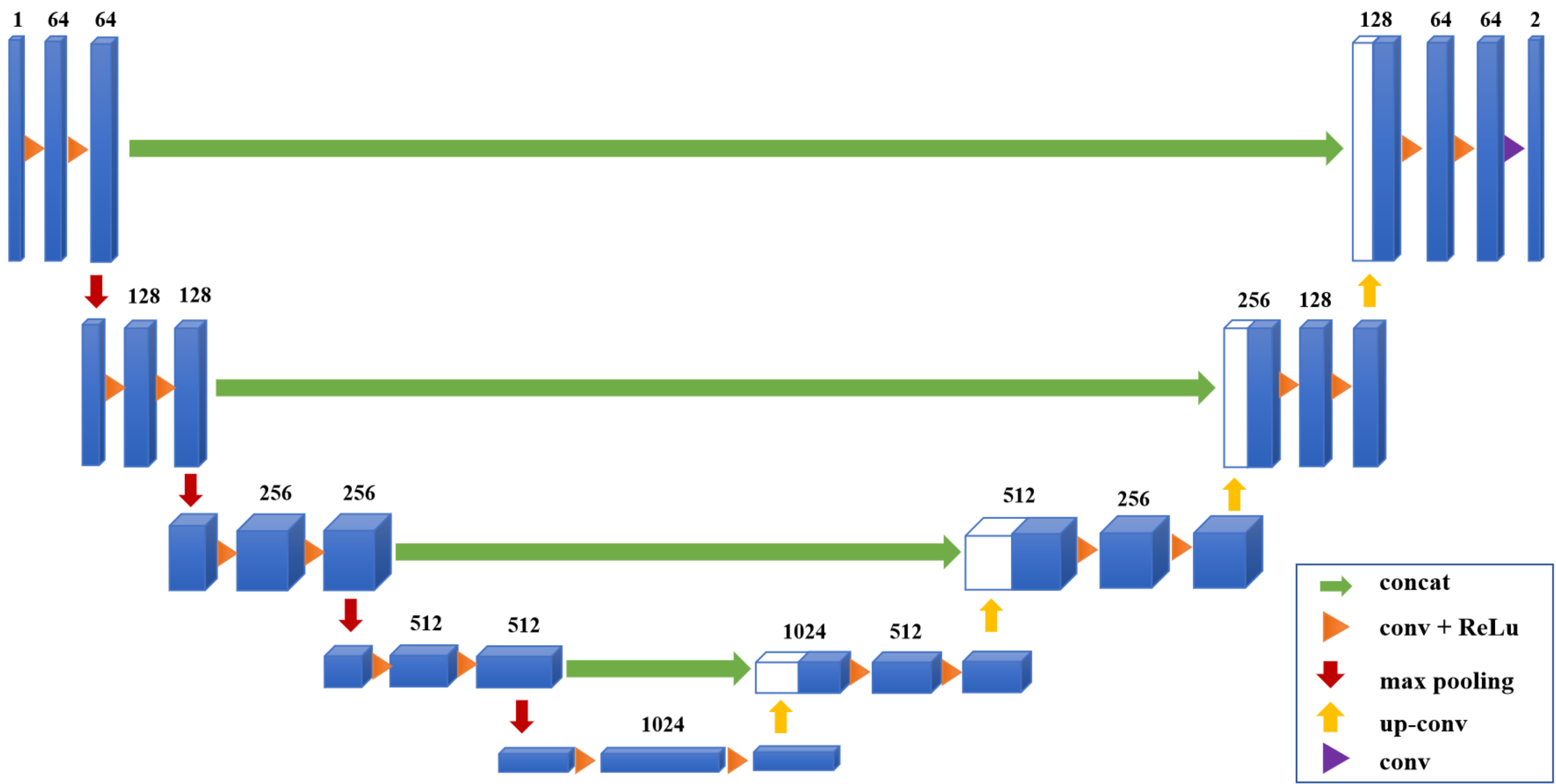}
  \caption{Representative U-net architecture for image segmentation task \cite{ref243}.}
  \label{figure:U-net}
\end{figure}

\textbf{U-net} is a popular image segmentation network which is developed primarily for medical image segmentation. It is built  based on \textbf{Convolutional Neural Network (CNN)} and is upgraded to achieve better segmentation performance \cite{ref120,ref208,ref207}. As shown in Figure \ref{figure:U-net}, the U-Net \cite{ref243} network is U-shaped, mainly including the downsampling layer on the left and the upsampling process on the right. The structure is symmetrical from left to right. Additionally, the encoder-decoder structure and skip connection used by U-Net are core designs to achieve a more effective combination of contextual information to make precise pixel-level judgments.

Alirr \cite{ref76} designs a lung image segmentation method to extract the region of interest, which modifies U-net with upgraded residual block including concatenation skip connection.   
Additionally, the segmented region employs edge enhancing diffusion filtering (EED) to improve the infection area's contrast. Their proposed method achieves Dice overlapping score of 96.1\% and 78.0\% for lung and infection areas segmentation. 
Diniz et al. \cite{ref65} seek to automatically identify infections caused by COVID-19 and evaluate quantitative score of the infected region. They propose a model by modifying the traditional U-net architecture via batch normalization, leaky ReLU, dropout, and residual block techniques. Their model yields an average Dice value of 77.1\% and an average specificity of 99.8\%.

Zhao et al. \cite{ref209} propose a novel dilated dual attention U-Net (D2A U-Net) with the dual attention strategy and hybrid dilated convolutions to segment COVID-19 lesion region. The former dual attention strategy, which contains two attention modules, is used to enhance feature maps and reduce the semantic gap among diverse levels of feature maps. The latter hybrid dilated convolutions are used for achieving larger receptive fields. Their proposed method achieves 0.73 Dice score and 0.71 recall score. 
Lian et al. \cite{ref210} propose an end-to-end DRD U-Net (dilated residual and deeply supervised U-Net) network to segment the lung lesions. They add residual modules to each layer of channel to accelerate convergence and avoid gradient disappearance. To extract richer feature information, the extended convolution unit is utilized to increase the receptive field without increasing the number of parameters. In the experiments, their proposed model shows impressive segmentation performance and the error can be significantly reduced compared with the original U-Net. 

U-net has shown great potential in the field of image segmentation. Some variants that follow the core idea of U-net have been proposed, including attention U-net \cite{ref212}, and Residual U-net \cite{ref213}, to complete more complex segmentation tasks. In addition to U-Net architecture, other segmentation methods such as V-Net, ResU-Net, Dense-Net, and DeepLab, are also used for the segmentation of COVID-19 infected region in different papers. Table \ref{table:segmentation_methods} displays the various segmentation techniques employed by different papers.
\begin{table}[htbp]
\centering
\caption{A summary of various segmentation methods used in different papers.}
\begin{tabular}{p{4cm}p{6cm}p{2cm}<{\centering}} 
\toprule
Segmentation methods\centering   & Papers\centering      & Count  \\
\midrule
U-Net\centering                  &\cite{ref63,ref64,ref65,ref66,ref69,ref72,ref73,ref75,ref76,ref77,ref78,ref79,ref80,ref81,ref9,ref242,ref210,ref211}\centering & 18     \\
V-Net\centering                  & \cite{ref69,ref241,ref192}\centering    & 3     \\
ResU-Net\centering               & \cite{ref65,ref72,ref73}\centering       & 3      \\
Dense-Net\centering              & \cite{ref68,ref78}\centering    & 2     \\
UNet++\centering                 & \cite{ref77,ref78} \centering   & 2      \\
Deeplab\centering                & \cite{ref65,ref67}\centering       & 2      \\
Attention-UNet\centering         & \cite{ref209,ref78} \centering      & 2      \\
SAUNet++ \centering              & \cite{ref66}\centering          & 1      \\
COVSeg-NET\centering             & \cite{ref71}\centering          & 1      \\
\bottomrule
\end{tabular}
\label{table:segmentation_methods}
\vspace{-0.5cm}
\end{table}

\subsection{Loss Function}
The loss function is another important component of the segmentation model. It can guide the segmentation model to learn towards a specific direction and segment the wanted infected area. Xiao et al. \cite{ref66} use focal Tversky loss (FTL) and the generalized Dice loss (GDL). The GDL can reduce the correlation between lesion size and Dice loss, and can effectively guide the model in segmenting small regions. Elharrouss et al. \cite{ref78} use binary cross entropy as loss function and their model reaches 78.6\% for Dice, 71.1\% for sensitivity metric, 99.3\% for specificity, and 85.6\% for precision. Alirr \cite{ref76} uses Dice loss as loss function and averages the DSC of each class to generate the final score. Diniz et al. \cite{ref65} use Dice loss as loss function and their proposed model reaches 77.1\% for Dice, and 99.76\% for average specificity. This section will introduce several representative loss functions commonly used in the medical image segmentation task.

\textbf{A. Binary Cross Entropy}

The pixel level cross entropy loss is commonly used in image segmentation task. For a given random variable, this loss function will examine each pixel individually and compare the difference between two probability distributions. Binary cross entropy is defined as \cite{ref163}:
\begin{equation}
L_{C E}(y, \hat{y})=-\left(y \log \left(\hat{y}\right)+\left(1-y\right) \log \left(1-\hat{y}\right)\right),
\label{CE}
\end{equation}

\noindent where $y$ and $\hat{y}$ represent the ground truth value and predicted value respectively.

\textbf{B. Weighted Binary Cross Entropy}

Weighted binary cross entropy (WCE) \cite{ref164} is a variant of binary cross entropy, which aims to solve the problem of uneven distribution of lesion information in medical image. WCE is defined as:
\begin{equation}
L_{W C E}(y, \hat{y})=-\left(\beta * y \log \left(\hat{y}\right)+\left(1-y\right) \log \left(1-\hat{y}\right)\right),
\label{WCE}
\end{equation}

\noindent where $\beta$ is used to tune false negative and false positive. If we want to decrease the number of false negative, set $\beta$>1. In contrast, setting $\beta$<1 can decrease the number of false positive.

\textbf{C.Balanced Cross Entropy}

Balanced cross entropy (BCE) \cite{ref165} is similar with WCE, which appropriately weights positive examples and negative examples. Balanced cross entropy is defined as:
\begin{equation}
L_{B C E}(y, \hat{y})=-\left(\beta * y \log \left(\hat{y}\right)+(1-\beta)\left(1-y\right) \log \left(1-\hat{y}\right)\right).
\label{BCE}
\end{equation}

\textbf{D. Focal Loss}

Focal loss \cite{ref166} is often used to deal with unbalanced sample classification. It focuses on weighting loss to the sample according to the difficulty of sample discrimination. For medical images presenting small lesion information, focal loss assigns a larger weight to positive pixels and can achieve excellent performance. It is defined as:
\begin{equation}
L_{f l}=-\alpha_t\left(1-p_t\right)^\gamma \log \left(p_t\right),
\label{FL}
\end{equation}

\noindent where $\gamma$ is a regulatory parameter and $\gamma$>0. $\alpha_t$ generally ranges from [0,1], which is used to deal with class imbalance. Focal loss defines $p_t$ as:
\begin{equation}
p_t= \begin{cases}\hat{p} & \text { if } y=1 \\ 1-\hat{p} & \text { otherwise }\end{cases},
\label{pt}
\end{equation}

\noindent where $\hat{p}$ represents the predicted value of pixel classification and $y$ denotes the ground truth value.

\textbf{E. Dice Loss}

Dice loss \cite{ref167} is designed to calculate the similarity between two images and is utilized to guide the training process of image segmentation methods.
Dice loss is defined as:
\begin{equation}
D L(y, \hat{p})=1-\frac{2 y \hat{p}+1}{y+\hat{p}+1}.
\label{DCS}
\end{equation}
Here, 1 is added in numerator and denominator to ensure that the function is not undefined in edge case scenarios where $y = \hat{p}= 0$.

\textbf{F. Generalized Dice Loss}

Dice loss is unsuitable to predict small targets. When some pixels of small targets are predicted incorrectly, Dice coefficient may greatly fluctuate. To solve these problems, GDL uses weights inversely proportional to lesion region, in order to better segment small regions \cite{ref66}. When the number of classes is 2, Generalized Dice loss is defined as:
\begin{equation}
G D L=1-2 \frac{\sum_{l=1}^2 w_l \sum_n r_{l n} p_{l_n}}{\sum_{l=1}^2 w_l \sum_n r_{l n}+p_{l_n}},
\label{GDL}
\end{equation}

\noindent where $r_{l n} \in\{0,1\}$ and $p_{l_n} \in[0,1]$ denote the true voxel values of $n$-th position and the related probability predicted as class $l$. $w_l$ represents the weight of class $l$ and is defined as:
\begin{equation}
w_l=\frac{1}{\left(\sum_{n=1}^N r_{l n}\right)^2+\varepsilon},
\label{wl}
\end{equation}

\noindent where $N$ is the total number of pixels, and $\varepsilon$ is often set as $10^{-5}$ to prevent the loss function from dividing by 0.

\textbf{G. Tversky Loss}

Tversky loss \cite{ref168} is considered as a generalization of Dice loss, where the coefficient $\beta$ is a hyperparameter that controls the balance between false negative (FN) and false positive (FP). Tversky loss is defined as:
\begin{equation}
T L(p, \hat{p})=\frac{p \hat{p}}{p \hat{p}+\beta(1-p) \hat{p}+(1-\beta) p(1-\hat{p})},
\label{TL}
\end{equation}

\noindent where $p$ and $\hat{p}$ denote the ground truth value and predicted value respectively. When $\beta$ = 0.5, it can be solved into regular Dice coefficient.

\textbf{H. Focal Tversky Loss}

Similar to focal loss, focal Tversky loss \cite{ref169} is proposed to deal with class imbalance. It focuses on learning hard samples with the help of  coefficient $\gamma$. Focal Tversky loss is often used to segment small regions and is defined as:
\begin{equation}
F T L=\sum_c\left(1-T I_c\right)^{\gamma},
\label{FTL}
\end{equation}

\noindent where the hyperparameter $\gamma$ can range from [1,3], and $T I_c$ (Tversky index) is defined as:

\begin{equation}
T I_c=\frac{\sum_{i=1}^N p_{i c} g_{i c}}{\sum_{i=1}^N p_{i c} g_{i c}+\alpha \sum_{i=1}^N p_{i \bar{c}} g_{i c}+\beta \sum_{i=1}^N p_{i c} g_{i \bar{c}}},
\label{TIc}
\end{equation}

\noindent where $p_{i c}$ and $p_{i \bar{c}}$ are probabilities that pixel $i$ belongs to the lesion class $c$ and non-lesion class $\bar{c}$, respectively. $g_{i c}$ and $g_{i \bar{c}}$ represent probabilities for another class. $N$ is the total number of pixels.
\vspace{+0.3cm}

\section{Overall Analysis of DL-based Methods for COVID‑19 Diagnosis}
Manual detection of COVID-19 (RT-PCR), which has extremely high accuracy and specificity, is always regarded as the gold standard for COVID-19 detection \cite{ref180,ref181}. However, RT-PCR is time-consuming, and the spread risk and susceptibility of the disease are greatly increased during detection \cite{ref145,ref147}. Hence, many studies are focusing on diagnosing COVID-19 using medical images, attempting to find a diagnostic method that can guarantee high accuracy and significantly minimize human contact during detection \cite{ref124,ref11,ref21,ref22}. Among the many inspection methods, the deep learning models using medical images have shown excellent performance. And they are thus extensively utilized in image classification, lesion segmentation, and severity quantification to fight against COVID-19, such as \textbf{Convolutional Neural Network (CNN)} \cite{ref26,ref99,ref37}, \textbf{Generative Adversarial Network (GAN)} \cite{ref103,ref151}, and \textbf{Long Short-Term Memory (LSTM)} \cite{ref109,ref182,ref108}. In this section, we will introduce the different applications of deep learning models in diagnosing COVID-19.

\subsection{CNN models for COVID-19 diagnosis}
\begin{figure}[htbp]
  \centering
  \includegraphics[width=\linewidth]{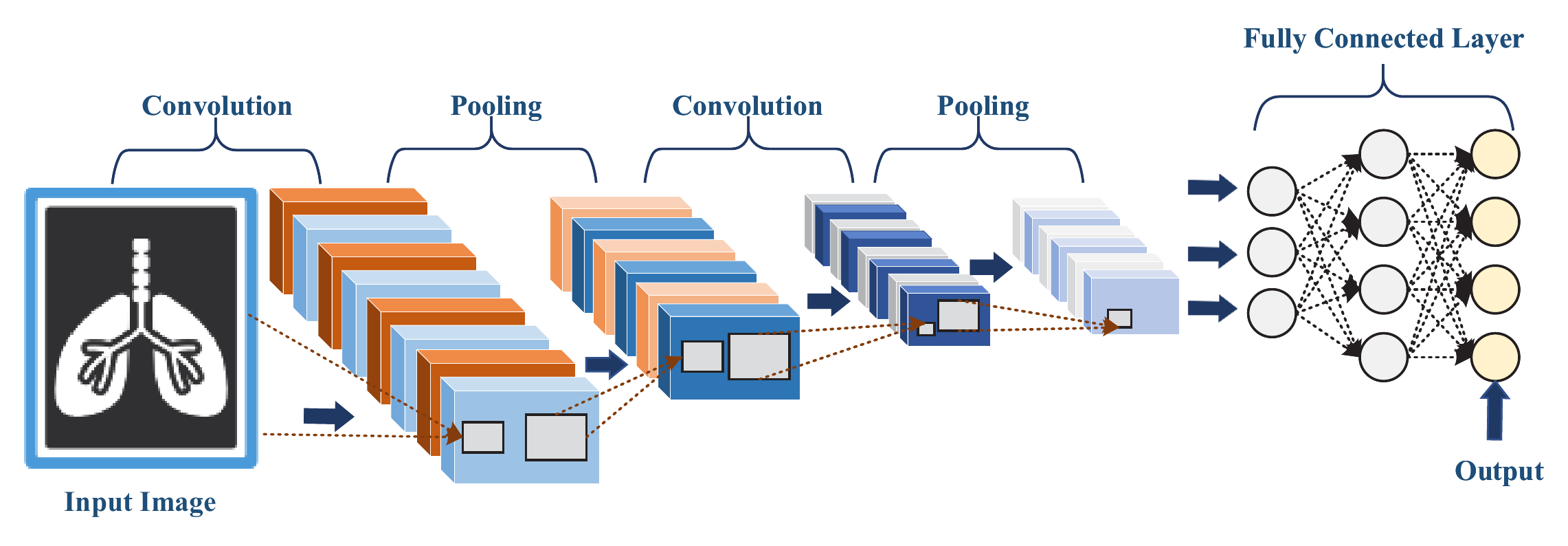}
  \caption{A representative workflow of using CNN model to identify COVID-19 cases from normal cases.}
  \label{figure:CNN_model}
  \vspace{-0.3cm}
\end{figure}

\textbf{Convolutional Neural Network (CNN)} is a multi-layer supervised neural network, in which the hidden convolutional and pooling layers are the core modules to realize the feature extraction function \cite{ref176,ref183}. CNN uses the gradient descent method to minimize the loss function and error back propagation (BP) is employed to reversely adjust the weight parameters in the network layer by layer \cite{ref184}. CNN models have demonstrated great potential in medical image processing and become a popular approach for identifying COVID-19 and localizing lesion. As Figure \ref{figure:CNN_model} depicts, a representative CNN model for COVID-19 diagnosis generally includes input, convolutional layers, pooling layers, fully connected layer, and output \cite{ref126}. Additionally, we try to record some deep learning methods and model evaluation metrics used in high quality studies and more detailed information is shown in Table \ref{table:CNN methods}.

Wang et al. \cite{ref51} design a deep learning-based method for automatic COVID-19 diagnosis. Their model uses a pre-trained U-Net to segment the lung region. The segmented 3D lung region is input into the 3D deep neural network to predict the possibility of infection. 
Finally, the activation regions in the classification network and the unsupervised connected components are together used to localize the COVID-19 lesions. Their proposed method obtains an accuracy of 0.90, a positive predictive value of 0.84, and a high negative predictive value of 0.98. Turkoglu \cite{ref48} proposes an expert-designed system named COVIDetectioNet, which is consisted of three basic components. First, deep features are obtained from the pre-trained AlexNet architecture. Second, the relief selection algorithm is employed to select the most efficient features from the pre-learned deep features. Third, the support vector machine (SVM) method is used for final classification. Their proposed model is validated by classifying 6,092 X-ray images as normal (healthy), COVID-19, and pneumonia cases, respectively. In the experimental results, their proposed model achieves 99.18\% accuracy and 97.1\% sensitivity.

Singh et al. \cite{ref26} propose an automated COVID-19 screening model. It employs some pre-trained CNN models such as DenseNet201, ResNet152V2, and VGG16, for early detection of COVID-19 patients. The predicted outputs are fed into the ensemble DCCNs, which are designed for diagnosing the suspected objects into four classes, including COVID-19 cases, tuberculosis, pneumonia, and healthy subjects. Their proposed model achieves 98.94\% accuracy, 98.94\% sensitivity and 98.93\% specificity, which is superior to the competitive models. Moreover, Aslan et al. \cite{ref214} firstly use ANN-based automated segmentation method to extract region of interest in the X-ray images. Second, various CNNs are used for feature extraction after data augmentation stage. Third, the features learned from each CNN model are input to four different machine learning (ML) algorithms for classification. The hyperparameters of each ML algorithm are determined by Bayesian optimization. Finally, the highest performance is obtained by the DenseNet201 model with SVM algorithm.

Kogilavani et al. \cite{ref218} explore the performance of a variety of convolutional neural network (CNN) models in detecting COVID-19, attempting to find the suitable deep models, including VGG16, DeseNet121, MobileNet, NASNet, Xception, and EfficientNet. Accuracy obtained for VGG16 is 97.68\%, DenseNet121 is 97.53\%, MobileNet is 96.38\%, NASNet is 89.51\%, Xception is 92.47\%, and EfficientNet is 80.19\%, respectively. Based on the performance analyses, it is demonstrated that the VGG16 architecture reaches the best accuracy compared to other architectures. Akbarimajd et al. \cite{ref216} propose a novel CNN approach applying adaptive convolution, which intends to enhance COVID-19 identification in noisy X-ray images without reducing noise. Specifically, the impulse noise-map layer and the adaptive resizing layer are added on the traditional CNN framework. A learning-to-augment strategy which uses noisy X-ray images is designed to improve the generalization of deep model. Some pre-trained networks such as SqueezeNet, ResNet18, and ResNet50 are modified to increase their robustness for impulse noise.
\begin{table}[h]
\centering
\caption{Deep learning methods and result evaluation for diagnosing COVID-19 using CNN models.}
\label{table:CNN result}
\resizebox{\textwidth}{!}{
\begin{tabular}{ccp{3.5cm}cccc}
\toprule
Reference & Data    & Deep Learning Methods\centering                             & Classification           & Acc & Sn & Sp \\
\midrule
Tang et al. \cite{ref98}        & X-ray  & Ensemble learning, COVID-Net                                              & Multiclass & 95.0\%     & 96.0\%     & \textbf{---}         \\
Yamaç et al. \cite{ref110}       & X-ray  & CheXNet, CSEN                                                             & Multiclass  & 95.9\%    & 98.5\%       & 95.7\%        \\
Pathak et al. \cite{ref54}        & CT     & CNN, Transfer learning                                                    & Binary     & 93.0\%   & 91.5\%      & 94.9\%       \\
Rathinasamy et al. \cite{ref111}       & X-ray  & CNN, Ensemble learning                                                    & Binary    & 99.0\%     &  \textbf{---}           & \textbf{---}           \\
Khan et al. \cite{ref112}       & X-ray  & COVID-RENet, SVM                                                          & Binary     & 98.5\%   & 99.0\%        &   \textbf{---}           \\
Wang et al. \cite{ref51}        & CT     & DeCoVNet, U-net                                                           & Binary    & 90.1\%    & 90.7\%       & 91.1\%        \\
Zhou et al. \cite{ref19}        & CT     & AlexNet, GoogleNet, ResNet, Ensemble learning                             & Multiclass  & 99.1\%   & 99.1\%      & 99.6\%        \\
Zheng et al. \cite{ref2}         & CT     & ResNet50, MAB, FAB                                                       & Binary    & 98.2\%   & 98.8\%      & 97.3\%       \\
Choudhary et al. \cite{ref8}         & CT     & VGG16, ResNet34                                                           & Binary     & 95.5\%  & 92.2\%      & 99.4\%       \\
Singh et al. \cite{ref26}        & CT     & CNN, Ensemble learning, Transfer learning                                  & Multiclass  & 98.8\%   & 98.8\%      & 98.8\%\\
Balaha et al. \cite{ref15}       &CT        &CNN, Transfer learning, GAN                                                &Binary & 98.7\% & \textbf{---}& \textbf{---} \\
Turkoglu \cite{ref48}           &X-ray      &AlexNet, Transfer learning, SVM                                             & Multiclass  &99.2\%  & 97.1\% &\textbf{---} \\
Aslan et al. \cite{ref214}      &X-ray & CNN, Transfer learning, Bayesian Optimization  & Multiclass & 96.3\% &96.4\% & 98.1\% \\
Zhang et al. \cite{ref217}      &CT & CNN, Transfer learning, Bayesian Optimization  &Binary & 92.1\% & \textbf{---} &91.2\%  \\
\bottomrule
\end{tabular}
\label{table:CNN methods}
}
\end{table}

Some existing convolutional neural network architectures have been proven to be vital in medical image feature extraction, such as ResNet \cite{ref85}, AlexNet \cite{ref86}, SqueezeNet \cite{ref87}, Inception \cite{ref88}, DenseNet \cite{ref89}, VGG \cite{ref90}, and EfficientNet \cite{ref91}. Various CNN models used in different papers for COVID-19 diagnosis are listed in Table \ref{table:CNN}. The ResNet is the most popular network model to detect COVID-19 and 34 papers use ResNet as backbone network. Besides, the DenseNet and the VGG are also used by many studies to detect COVID-19. In conclusion, it is significant for us to select an appropriate CNN model as backbone network to diagnose COVID-19 by considering data type, actual scale, application scenarios and other factors.

\subsection{Transfer learning for COVID-19 diagnosis}
\begin{figure}[h]
  \centering
  \includegraphics[width=\linewidth]{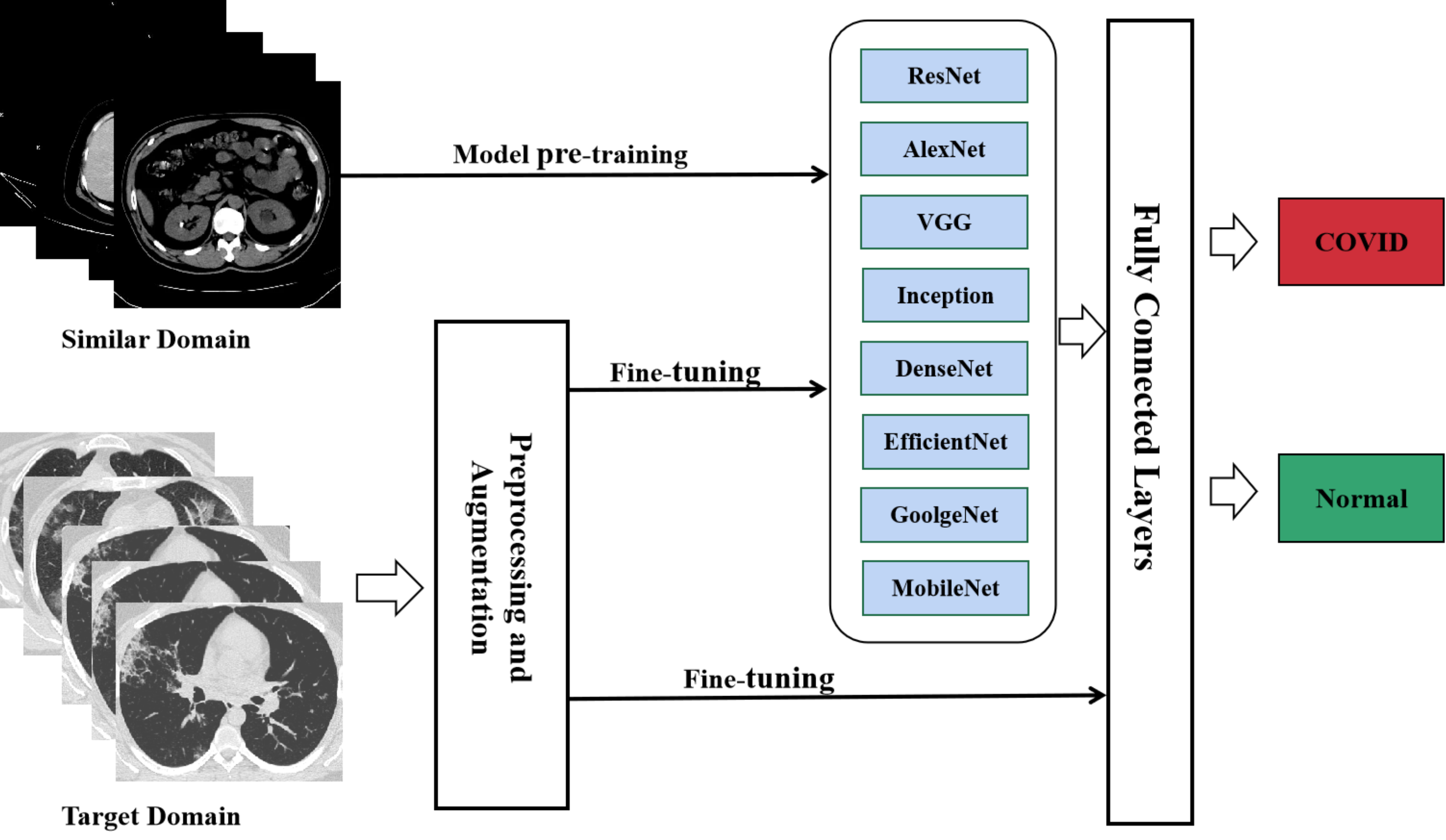}
  \caption{The application of deep learning using transfer learning method for COVID-19 diagnosis.}
  \label{figure:transfer learning}
\end{figure}

\textbf{Transfer learning} is a machine learning technique that can transfer existing knowledge from one domain (source domain) to another (target domain) \cite{ref186}. The source domain generally has sufficient annotated data and many existing models have learned excellent feature extraction capabilities. The target domain lacks large-scale annotated samples and the cost of obtaining annotated samples is too high \cite{ref187,ref188}. Transfer learning aims to use knowledge learned from the source domain to help the target learner achieve better performance. The closer the relationship between the target domain and the source domain, the better transfer learning can be achieved \cite{ref128}. Otherwise, it may be more difficult and even has negative transfer to bring harmful effects. As shown in Figure \ref{figure:transfer learning}, the pre-trained model has obtained excellent generalization performance in the source domain and then it is fine-tuned by using small-scale data from the target domain \cite{ref149}. Due to the expensive cost of capturing CT or X-ray images of COVID-19 patients, various pre-trained deep learning models have been employed for COVID-19 diagnosis.

Kabe et al. \cite{ref93} propose a novel domain transfer learning model for classifying COVID-19 cases, named feature fusion, decompose and transfer (FFDT). Their proposed FFDT gains feature enhancement by combining far-off features taken from far-off domains into a single feature space where the distribution mismatch is reduced. 
Additionally, they adopt modified convolutional neural network (MCNN) to extract features and the class reconstruction is used to unravel the local structure of the data distribution. Their model achieves the classification accuracy of 94.5\%. Lu et al. \cite{ref95} suggest that transfer learning can be used to extract features from chest CT images because it is of high complexity to train a CNN model from scratch. The pre-trained ResNet-18 and ResNet-50 models are chosen as the backbone to extract features from the CT images. To create refined image features, the retrieved features are combined using discriminant correlation analysis. Finally, in order to get a more reliable classification performance, three randomized neural networks are trained using the improved features and their predictions are combined. 

Vogado et al. \cite{ref94} evaluate transfer learning techniques in five pre-trained CNN architectures (VGG-16, VGG-19, ResNet-50, Xception and DenseNet-121). In specific, they train CNNs using 1,436 images associated with COVID-19 cases, 1,932 healthy images, and 3,651 images of other pathologies. The pre-trained ResNet50 obtains the best performance for extracting deep features, and the MLP classifier has shown the best results when using the features derived by ResNet50. Makris et al. \cite{ref92} use a combination of publicly available X-ray images from patients with confirmed COVID-19 cases, common bacterial pneumonia and healthy cases. Transfer learning is used for mitigating the issue of insufficient samples. In order to classify images, VGG16, VGG19, MobileNet V2, Inception V3, Xception, InceptionResNet V2, DenseNet201, and ResNet152 V2 are employed and compared to explore the best model for COVID-19 diagnosis. Specifically, VGG16 and VGG19 perform the best and achieve an overall accuracy of 95\%.

Pathak et al. \cite{ref54} employ a deep transfer learning technique to classify COVID-19 infected patients. The ResNet-50 is chosen as backbone network and transfer learning is employed to adjust the initial parameters of the deep layers. The pre-trained model, which is fine-tuned by a small number of training samples, can extract correct deep features from chest CT images to detect COVID-19 infected patients. Their proposed model achieves training and testing accuracy up to 96.23\% and 93.02\% respectively, which is superior to the compared deep models. 
To detect COVID-19 by medical images, Shamsi et al. \cite{ref219} propose a novel deep uncertainty-aware transfer learning framework. 
The pre-trained CNNs are used to extract features from chest X-ray and CT images. The extracted features are subsequently used to identify COVID-19 cases using various machine learning and statistical modeling techniques. It has been found that CT images can produce superior diagnosis because they contain more information than X-ray images.
\begin{table}[h]
\centering
\caption{A summary of various CNN models used by different papers for COVID-19 detection.}
\resizebox{\textwidth}{!}{
\begin{tabular}{cp{10cm}c} 
\toprule
CNN model  & Papers\centering  & Count  \\
\midrule
ResNet & \cite{ref2,ref4,ref9,ref10,ref13,ref15,ref18,ref19,ref20,ref22,ref24,ref26,ref29,ref31,ref35,ref36,ref40,ref41,ref45,ref47,ref48,ref49,ref50,ref51,ref53,ref54,ref55,ref56,ref57,ref58,ref60,ref61,ref216,ref214}             & 34   \\
DenseNet     & \cite{ref9,ref13,ref18,ref22,ref24,ref26,ref27,ref34,ref41,ref50,ref55,ref60,ref61,ref59,ref214,ref218}         & 16     \\
VGG          & \cite{ref4,ref15,ref24,ref26,ref31,ref39,ref41,ref42,ref48,ref49,ref55,ref56,ref57,ref61,ref62,ref218}         & 16     \\
MobileNet    & \cite{ref10,ref12,ref15,ref27,ref31,ref38,ref52,ref216,ref214,ref218,ref215,ref217} & 12      \\
Inception    & \cite{ref9,ref15,ref24,ref31,ref47,ref48,ref50,ref55,ref60,ref61,ref214}                 & 11     \\ 
AlexNet      & \cite{ref7,ref19,ref21,ref50,ref51,ref52,ref56,ref214}          & 8      \\
GoolgeNet    & \cite{ref18,ref19,ref50,ref53,ref56,ref216,ref214}           & 7      \\
Xception     & \cite{ref15,ref22,ref31,ref38,ref47,ref218}       & 6     \\
EfficientNet & \cite{ref4,ref15,ref28,ref38,ref57,ref218}              & 6     \\
SqueezeNet   & \cite{ref9,ref18,ref216}          & 3      \\
Darknet      & \cite{ref18,ref38}       & 2      \\
\bottomrule
\end{tabular}
}
\label{table:CNN}
\vspace{-0.3cm}
\end{table}

\subsection{Ensemble learning for COVID-19 diagnosis}
\begin{figure}[h]
  \centering
  \includegraphics[width=\linewidth]{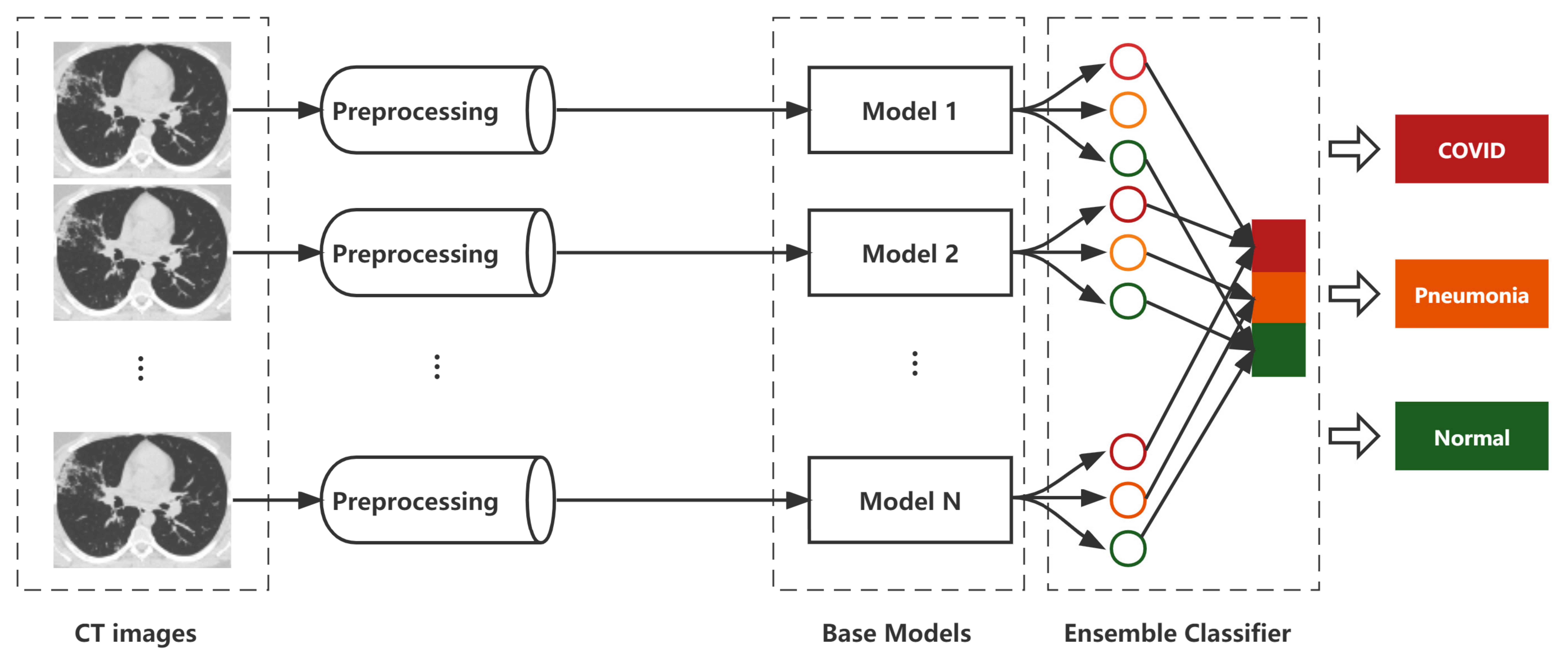}
  \caption{The application of deep learning using ensemble learning method for COVID-19 diagnosis.}
  \label{figure:ensemble learning}
  \vspace{-0.3cm}
\end{figure}

\textbf{Ensemble learning} is a machine learning method that combines multiple learners to complete complex learning tasks \cite{ref189,ref190}. It trains multiple base learners and combines them to obtain improved generalization ability than the individual base learners \cite{ref96}. Currently, the common methods to generate base learners can be divided into two categories: one is to apply different types of learning models in the same size but different samples, which are generated from the same dataset \cite{ref191}. The base learners which are generated by this method are called heterogeneous learners. The other is to apply the same learning model on different training sets. The base learners which are generated by this method are called homogeneous learners. Besides, the combined strategy of base learners mainly includes simple average method, weighted average method, majority voting, plurality voting, and weighted voting. Figure \ref{figure:ensemble learning} illustrates a representative application of ensemble learning for COVID-19 diagnosis. Multiple base models learn to diagnose COVID-19 from the medical images and output classification results. The ensemble classifier receives input from base models to make the final decision.

Chaudhary et al. \cite{ref97} train three base models (two Efficient-Net with different initial pre-trained weights and SE-ResNext), which can classify X-ray images into COVID-19, pneumonia, and normal cases. The final results are calculated by averaging the classification outcomes produced by the three different models individually. Their proposed method achieves excellent performance with an accuracy of 0.9592, a sensitivity of 0.9592, and a specificity of 0.9597. It is proven that the ensemble model's accuracy is greater than that of three separately trained models. Tang et al. \cite{ref98} propose the EDL-COVID model using deep learning and ensemble learning. It trains several base models to overcome the shortcomings of single model by combining their predicted outputs. A deep learning network generates several model snapshots in one training run by snapshotting. Additionally, several model snapshots are integrated to produce a preciser model and the final classification is made using the weighted average method. Their EDL-COVID model achieves 95\% accuracy. 

Abraham et al. \cite{ref38} use five pre-trained CNNs (MobilenetV2, Shufflenet, Xception, Darknet53, and EfficientnetB0) to extract features. Then the features are combined, and the ensemble classifier kernel support vector machine is used to diagnose COVID-19 cases. Their proposed model achieves 0.916 accuracy, 0.8305 kappa score, 0.91 F-score, 0.917 sensitivity, and positive predictive value of 0.904. Zhou et al. \cite{ref19} employ three deep pre-trained models (AlexNet, GoogleNet, and ResNet) as base learners. Then the predicted outputs from three pre-trained models are input into the ensemble classifier EDL-COVID, and relative majority voting is used to determine the final outcome. Finally, by comparing the ensemble classifier with three component classifiers in some specific performance metrics, it is demonstrated that the ensemble model obtains more effective performance than three deep pre-trained models. 

\subsection{GAN model for COVID-19 diagnosis}
\begin{figure}[htbp]
  \centering
  \vspace{-0.3cm}
  \includegraphics[width=\linewidth]{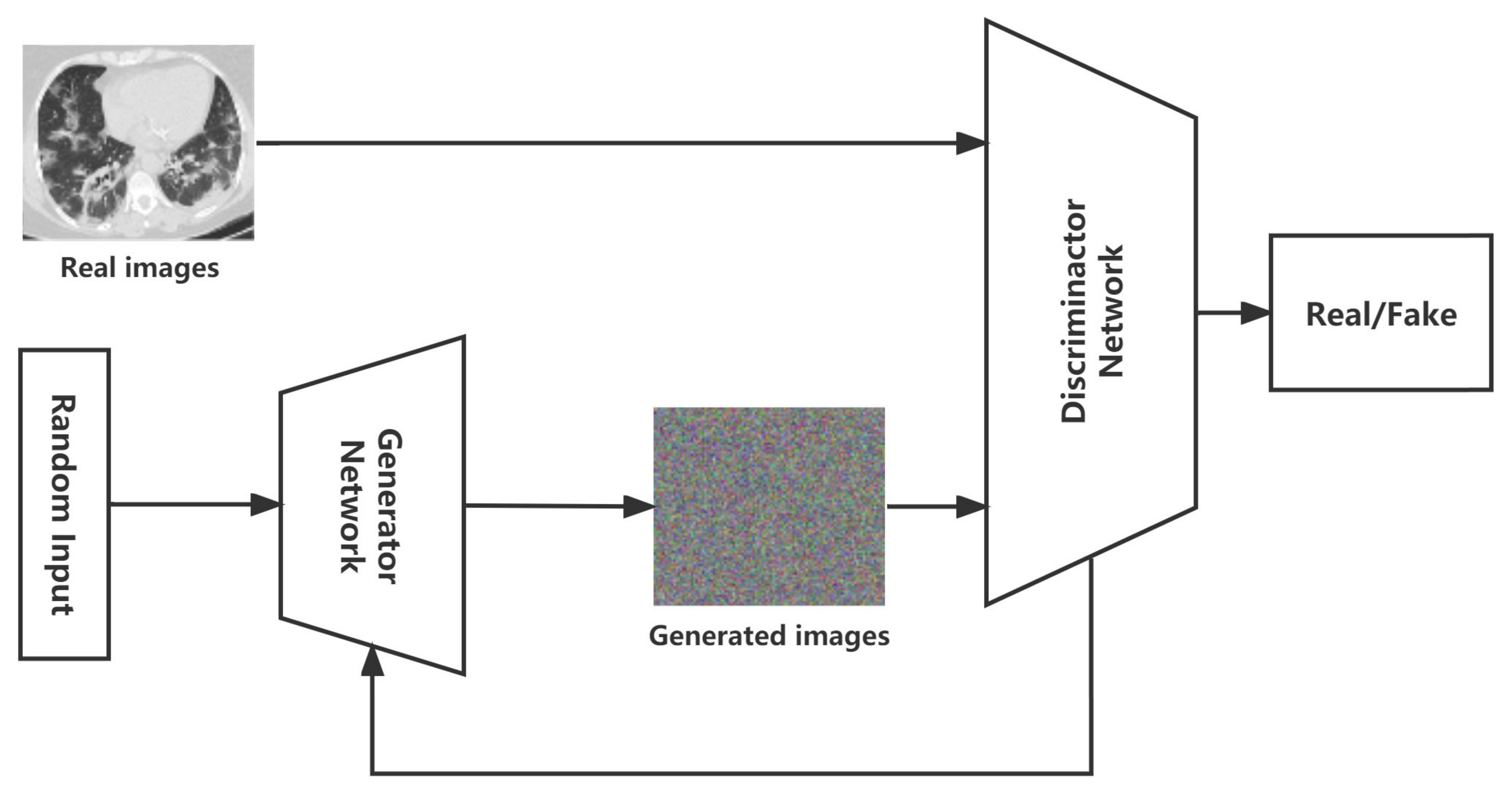}
  \caption{The typical architecture of the GAN. The generator generates synthetic data from a given input, the discriminator distinguishes the output of the generator from the real data.}
  \label{figure:GAN}
\end{figure}

One of the main causes for poor performance and underfitting is the insufficient amount of annotated COVID-19 images \cite{ref145,ref149}. And the cost of collecting COVID-19 infected images is too expensive. To solve these problems, \textbf{Generative Adversarial Network (GAN)} is employed by many studies to generate fake COVID-19 infected images to tackle data insufficiency \cite{ref196,ref101}. GAN mainly consists of two systems: a generator and a discriminator, as shown in Figure \ref{figure:GAN}.

\begin{itemize}
\item[$\bullet$]
\textbf{Generator network:} Generator takes noise samples from a particular distribution (uniform distribution and Gaussian distribution) and generates results similar to the real training data. It tries to generate fake images to successfully trick the discriminator after training.
\end{itemize}

\begin{itemize}
\item[$\bullet$]
\textbf{Discriminator network:} Real data and generated data are mixed and input into the discriminator. The discriminator distinguishes whether a sample belongs to real data or generated data. If it comes from real data, then the high probability will appear to be output, otherwise the probability is low.
\end{itemize}

Jiang et al. \cite{ref100} construct a public COVID-19 CT dataset, including 1,186 CT images synthesized from a large-scale lung cancer CT dataset using CycleGAN. Their proposed model can learn the GGO style of COVID-19 so that the synthetic images are closely resembled to the real distribution. Goel et al. \cite{ref102} employ generative adversarial network (GAN) to generate synthetic chest CT images during data augmentation phase. The Whale Optimization Algorithm (WOA) is used to optimize the hyperparameters of GAN. Their proposed model using GAN reaches 99.22\% accuracy.

Bargshady et al. \cite{ref103} apply generative adversarial network (GAN) and semi-supervised CycleGAN (SSA-CycleGAN) to augment the training dataset of X-ray images. Their proposed Inception-CycleGAN model achieves 94.2\% accuracy, and 92.2\% area under cure. Serte et al. \cite{ref104} augment the number of available CT images by using generative adversarial network (GAN). By comparing the traditional deep learning methods with their proposed method using data-efficient method (GAN), it is demonstrated that their proposed data-efficient model outperforms all other traditional deep learning models. The ResNet-18 and MobileNetV2 obtain the best performance.

In addition to its popularity in data augmentation, GAN is also employed in other fields based on its adversarial training characteristics. Bhattacharyya et al. \cite{ref105} use C-GAN to segment the COVID-19 chest X-ray images. The X-ray images as input are fed into the generator and the generator network tries to produce the mask images. The discriminator tries to distinguish fake image pair (Input X-ray images and generated mask images) from the real one. Besides, Doraiswami et al. \cite{ref106} propose an effective prediction mechanism, where local ternary pattern (LTP) is used for feature extraction, and prediction of COVID-19 patients is performed by their proposed Jaya-TSA based GAN after acquiring features. Their proposed method reaches the accuracy of 0.87, the sensitivity of 0.85 and the specificity of 0.89.

Some of the applications of deep learning based on GAN model and their result are demonstrated in Table \ref{table:GAN result}. In most of the studies, GAN is generally employed for data augmentation before model training to greatly increase the training sample space, thereby significantly improve the performance of the model \cite{ref194,ref198,ref104}. In reference \cite{ref194}, the quantitative analysis shows that their proposed MTT-GAN greatly improves the accuracy of binary classifier and multiclass classifier. In reference \cite{ref198}, the original CNN models with Generative Adversarial Network (GAN) increase the accuracy by 2\% to 3\%, the recall by 2\% to 4\%, and the precision by 1\% to 3\%.
\begin{table}[h]
\centering
\caption{Deep learning methods and result evaluation for diagnosing COVID-19 using GAN models.}
\label{table:GAN result}
\resizebox{\textwidth}{!}{
\begin{tabular}{ccp{3.5cm}cccc}
\toprule
Reference & Data    & Methods\centering    & Function          & Acc & Sn & Sp    \\
\midrule
Zhang et al. \cite{ref113}       & CT     & GAN, U-net         &Segmentation &93.2\% &69.8\% & \textbf{---} \\  
Bargshady et al. \cite{ref103}       & X-ray  & CycleGAN, Inception V3  & Classification  &94.2\% &95.5\% &91.4\% \\  
Acar et al. \cite{ref101}       & CT & GAN,CNN                   & Classification &95.0\%& 94.2\% &95.3\%  \\
Goel et al. \cite{ref102}       & CT & GAN, WOA, ResNet-50        & Classification  & 99.2\% & 99.8\% & 97.8\% \\                                      
Menon et al. \cite{ref194}       & X-ray  & GAN, CNN, transfer learning  & Classification    &96.3\% & 100.0\% &93.2\% \\                            
Li et al. \cite{ref198}       & CT  & GAN, DenseNet              & Classification &93.0\% &96.0\% &\textbf{---} \\
Serte et al. \cite{ref104}    & CT &GAN, CNN, Transfer learning  & Classification &74.0\% & 88.0\% &68.0\%     \\
Bhattacharyya et al. \cite{ref105} & X-ray &C-GAN, CNN, ML  & Segmentation &96.6\% &95.0\% &97.4\%            \\
\bottomrule  
\end{tabular}
}
\end{table}

\subsection{LSTM model for COVID-19 diagnosis}
\textbf{Long Short-Term Memory (LSTM)} is a powerful recurrent neural network, which is proposed to address the substantial limitation of neural networks when dealing with sequential data \cite{ref107}. For the issue that traditional RNN structure has poor performance on long sequences, the LSTM model is capable of greatly alleviating the gradient disappearance problem, thereby supporting long-term dependence in dealing with sequential data \cite{ref197,ref195}. And it has been extensively utilized to diagnose COVID-19 and predict the prognosis of COVID-19 patients. In this section, we will introduce the application of LSTM model in detecting COVID-19, and some high quality studies using the LSTM model for
COVID-19 diagnosis are listed in Table \ref{table:LSTM result}.
\begin{figure}[htbp]
  \centering
  \includegraphics[width=\linewidth]{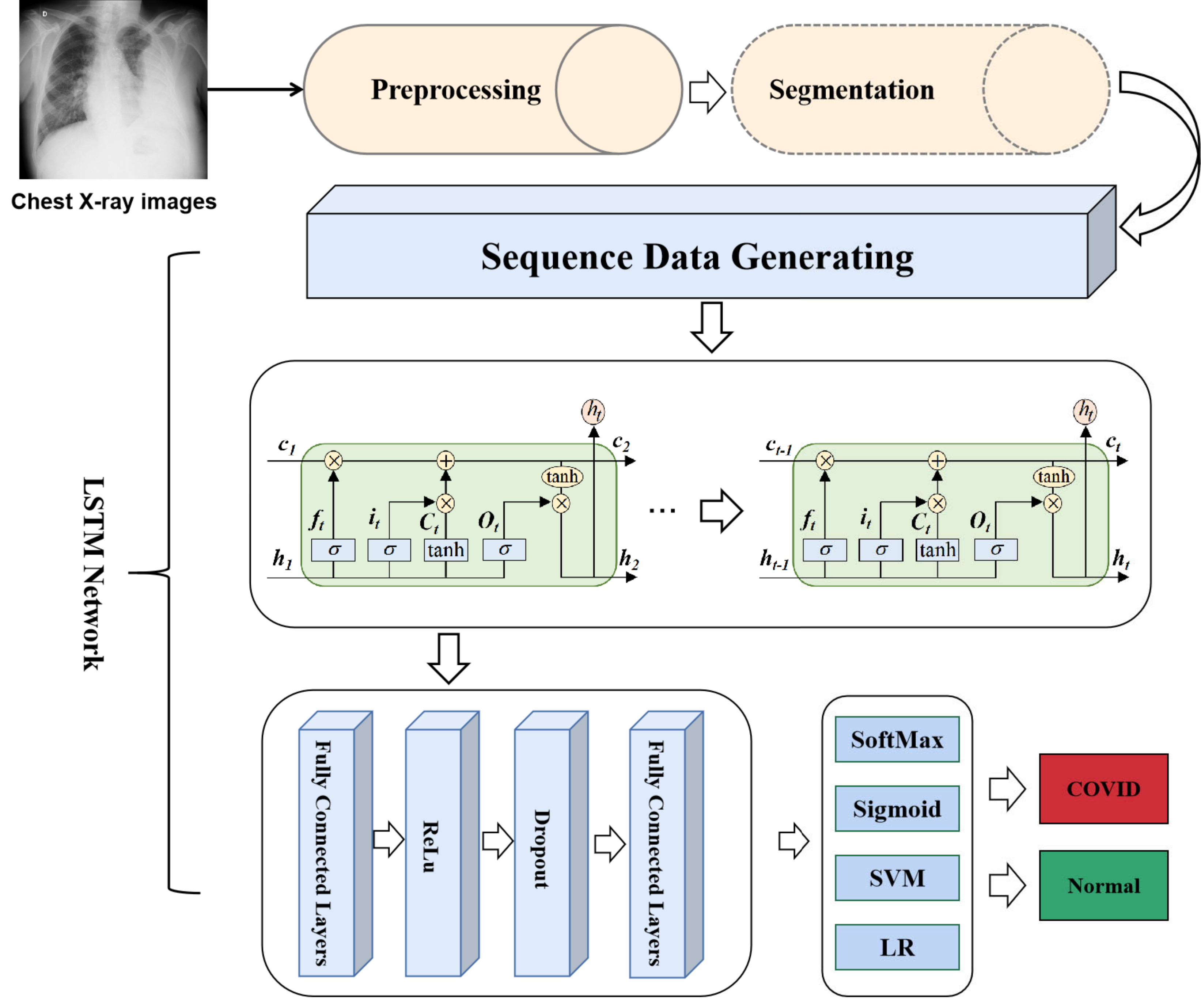}
  \caption{The application of deep learning using LSTM model for COVID-19 diagnosis.}
  \label{figure:LSTM}
\end{figure}

Patients with COVID-19 tend to have a dynamic condition. The information obtained from one CT image or X-ray is generally limited. Conversely, CT or X-ray sequences are capable of providing more medical information, assisting models or doctors to make a preciser diagnosis for COVID-19 patients \cite{ref11}. Consequently, many studies apply the LSTM model and image sequences to diagnose COVID-19. The typical architecture of the LSTM model for COVID-19 diagnosis is shown in Figure \ref{figure:LSTM} \cite{ref109}. Er \cite{ref220} proposes a hybrid approach for COVID-19 classification that combines long short-term memory (LSTM) with some pre-trained deep networks. The contrast enhancing method is firstly applied to X-ray images in preprocessing phase. Then the pre-trained CNNs and LSTM model are used to learn features from the contrast enhanced chest X-rays. Finally, COVID-19, normal (healthy), and pneumonia cases are classified by softmax. Their proposed model reaches 98.97\% accuracy, 98.80\% precision, and 98.70\% sensitivity.

Hasan et al. \cite{ref107} preprocess the CT images to reduce the effect of intensity variations. Then, histogram thresholding is used to segment the CT lung region. Each CT image is extracted feature using Q-Deformed entropy (QDE) and convolutional neural network (CNN). Then the obtained features are fused and fed into a long short-term memory (LSTM) classifier to identify COVID-19 cases. The highest accuracy for classifying the collected dataset is 99.68\%. Sheykhivand et al. \cite{ref108} propose an efficient deep neural network for COVID-19 automatic detection. First, the GAN model is used to generate sufficient train samples for augmenting data. Later, the pre-trained model (Inception) obtains a feature vector from the X-ray images. Then the feature vector is split into shorter vector sequences for the input of LSTM model to identify COVID-19. Their proposed model achieves more than 90\% accuracy for most scenarios and the accuracy of 99\% for separating COVID-19 from healthy group.

Xu et al. \cite{ref11} develop a three dimensional algorithm that combines multi-instance learning with the LSTM architecture (3DMTM) to identify COVID-19 from community acquired pneumonia (CAP). The 3DMTM model employs a lesion instance generator based on a pneumonia segmentation model to generate many lesion instances, which are then combined with clinical information and input into LSTM for final classification. Their proposed model achieves 0.956 AUC, 0.862 and 0.98 specificity under the condition of relatively large data. 
Demir \cite{ref109} proposes a novel LSTM-based model to automatically identify COVID-19 cases using X-ray images. Their proposed model is trained from scratch. In particular, the sobel gradient and marker-controlled watershed segmentation operations are firstly applied to raw images. The processed images are then converted to sequence data, which is fed into LSTM layer to generate the feature vector. The feature vector is finally input into softmax layer to identify COVID-19 cases. In the experiment, regarding the accuracy, sensitivity, specificity, and F-score, the proposed method achieves excellent performance.

\section{Quantify the Severity of COVID-19 Patients}
Early identifying COVID-19 from the patient's CT images or X-ray images is the first and crucial step in COVID-19 diagnosis. After that, a large number of confirmed and suspected cases need to be managed properly and given appropriate treatment, which is an enormous challenge to medical resource distribution \cite{ref118,ref5,ref115}. To solve these problems, quantifying the severity of COVID-19 patients is another crucial component for COVID-19 diagnosis, which can help doctors have a better grasp of the patient's infected condition and make the most reasonable treatment, thereby maximizing the rational distribution of medical resources.

Rana et al. \cite{ref114} design a severity estimation SSD network, which uses the images collected from the detection experiment as the training set. Based on the COVID-19 positive images, their proposed model predicts different feature classes and bounding box coordinates. Then the evaluation module uses the top 36 predicted classes, ignoring the background classes, to derive severity estimates. In the end, three classifications are created based on the severity of COVID-19 patients, mainly including initial type, intermediate type, and severe type.

Shan et al. \cite{ref81} develop a DL-based segmentation method using VB-Net to segment COVID-19 infection regions in CT images. Each training case's automatic annotation is improved using the suggested HIMI (human involved model iterations). The chest CT scans are first fed into the proposed segmentation model. Then, quantitative metrics, e.g., infection volumes and POIs in the entire lung, lung lobes, and bronchopulmonary segments, are estimated to characterize the infection locations in the CT image.

Chamberlin et al. \cite{ref116} try to evaluate a previously trained interpretable deep learning algorithm for the diagnosis and prognosis of COVID-19. Three radiologists with cardiothoracic fellowship training systematically evaluate each chest radiograph and generated a severity score based on the region of airspace disease. And the evaluation results are compared with the identical score produced by artificial intelligence. It shows that the anticipated severity scores are compatible with professional evaluation and AI model correctly forecasts crucial clinical consequences.

Zhou et al. \cite{ref221} propose a multi-modality feature learning and fusion model for COVID-19 patient severity prediction. The CT images and electronic medical record (EMR) are used for multi-modality feature extraction. The High-order Factorization Network (HoFN) is proposed to learn the impact of a set of clinical features from an electronic medical record (EMR). Finally, the features are concatenated as the input of fully connected layer to evaluate a patient’s severity. In general, according to clinical symptoms and medical image data, their proposed model classifies patients' severity as mild, moderate, severe, or fatal.
\begin{table}[htbp]
\centering
\caption{Deep learning methods and result evaluation for diagnosing COVID-19 using LSTM models.}
\label{table:LSTM result}
\begin{tabular}{ccp{3cm}ccc}
\toprule
Reference & Data   & Methods\centering  & Acc & Sn & Sp   \\
\midrule
Naeem et al. \cite{ref182}       & CT and X-ray & CNN, LSTM   &98.9\% &99.0\% & \textbf{---}    \\
Aslan et al. \cite{ref199}       & X-ray  & AlexNet, LSTM    & 98.7\% &98.8\% &99.3\%            \\
Hamza et al. \cite{ref200}       & X-ray               & Efficient Net, LSTM  & 93.4\% & 93.3\% & \textbf{---} \\
Demir \cite{ref109}       & X-ray               & LSTM      &97.6\% &100.0\% &96.0\% \\
Sheykhivand et al. \cite{ref108}       & X-ray              & CNNs, GAN, LSTM   & 99.5\% &100.0\% & 99.0\%  \\
Hasan et al. \cite{ref107}       & CT &LSTM, Q-Deformed Entropy  & 99.7\% &\textbf{---} &\textbf{---}   \\
Xu et al. \cite{ref11} & X-ray & LSTM, SVM & 95.3\% &86.2\% &98.0\% \\
Er \cite{ref220} &X-ray & CNN,LSTM & 99.0\%  &98.7\% & \textbf{---} \\
\bottomrule
\end{tabular}
\end{table}

\section{Challenges and Future Work}
Deep learning technologies have been proven to have great potential in fighting against COVID-19 and are widely used for diagnosis and quantification. However, the applications of deep learning for COVID-19 diagnosis are still in their infancy with many shortcomings. In this section, we will detail the challenges and future work when applying deep learning technologies to diagnose COVID-19.

\subsection{Challenges}
At present, the applications of deep learning based on medical images for diagnosing COVID-19 mainly face eight challenges:
\begin{itemize}
\item[$\bullet$]
\textbf{Accuracy in multiclass classification.} Although the deep learning model based on medical images can identify COVID-19 from normal cases, it is still challenging to distinguish COVID-19 through different types of pneumonia, which is far less accurate than RT-PCR.
\end{itemize}

\begin{itemize}
\item[$\bullet$]
\textbf{Unavailability of large-scale annotated data.} The performance of most deep learning methods depends on large-scale annotated data. Although some studies propose their own datasets, the available data is still insufficient. Additionally, annotating data is time-consuming and requires many professional medical people.
\end{itemize}

\begin{itemize}
\item[$\bullet$]
\textbf{Data imbalance.} Due to the rapid outbreak of the epidemic, the positive COVID-19 samples are far smaller than the normal samples. This data imbalance will affect the performance of deep learning models in COVID-19 diagnosis.
\end{itemize}

\begin{itemize}
\item[$\bullet$]
\textbf{Image quality.} Artifacts and noise often appear in datasets or imaging methods or in the devices used to capture images, which may interfere with the learning direction of the deep models to make incorrect judgments. Therefore, more effective noise reduction methods and dataset cleaning technologies need to be studied.
\end{itemize}

\begin{itemize}
\item[$\bullet$]
\textbf{Lack of multi-modality based system.} Most studies employ only one of medical images (CT or X-ray images) to diagnose COVID-19, which is insufficient in complex infection situations. A multi-modality based system can take the advantages of all modalities and meet more complex requirement of diagnosis.
\end{itemize}

\begin{itemize}
\item[$\bullet$]
\textbf{Lack of research on 3D images.} With the development of medical imaging, 3D images have been used with richer medical information. However, most advanced deep learning models are trained on 2D images, which may ignore many important features.
\end{itemize}

\begin{itemize}
\item[$\bullet$]
\textbf{The intersection of computer science and medicine fields.} The applications of deep learning in the fight against COVID-19 require deep collaboration in computer science, medical imaging, bioinformatics, virology and many other related fields.
\end{itemize}

\begin{itemize}
\item[$\bullet$]
\textbf{Data privacy.} Facing the outbreak of COVID-19, most studies require a series of data such as personal information, image, and clinical record of patients. A question worth thinking is how to effectively protect the privacy and human rights of patients in the fight against COVID-19 based on deep learning.
\end{itemize}

\subsection{Future Work}
In order to make better use of medical images and deep learning to fight against COVID-19, future studies can study the COVID-19 diagnosis from the following different perspectives.

\begin{itemize}
\item[$\bullet$]
\textbf{Multiclass classification.} To better fight against COVID-19 and meet the complex infection situation, the future diagnostic model should consider the direction of multiclass classification. Only if the deep learning technology has a very high accuracy in multiclass classification, can it possibly outperform the existing RT-PCR.
\end{itemize}

\begin{itemize}
\item[$\bullet$]
\textbf{Non-contact detection.} During CXR and CT image inspection, the non-contact image acquisition can significantly decrease the infection risk between radiologists and patients in the COVID-19 pandemic.
\end{itemize}

\begin{itemize}
\item[$\bullet$]
\textbf{Data fusion.} Numerous annotated samples are required by deep learning models and different hospitals and institutions, which have different data collection protocols. The types and formats of collected data can be significantly different from each other. Hence, a very meaningful direction is to use data fusion methods to build a large-scale dataset.
\end{itemize}

\begin{itemize}
\item[$\bullet$]
\textbf{Transfer learning.} Most studies employ a small-scale dataset, which is likely to lead to model underfitting and poor performance. Transfer learning is worth studying to speed up the training of models and improve the performance by transferring knowledge from similar domains.
\end{itemize}

\begin{itemize}
\item[$\bullet$]
\textbf{Incremental learning.} In almost all areas of the fight against COVID-19, the dataset and new studies are growing steadily. One question worth noticing is how to improve the ability to optimize old knowledge while absorbing new knowledge. Thus we suggest all the models should be implemented in an incremental way.
\end{itemize}

\begin{itemize}
\item[$\bullet$]
\textbf{Multi-modality.} A single-modality based prediction systems generally have limited predictive performance. In the future, we can consider integrating multiple image information, and even combine with clinical information, which will greatly improve the performance of model and interpretability.
\end{itemize}

\begin{itemize}
\item[$\bullet$]
\textbf{Optimization.} Many problems in deep learning are optimization problems, and existing methods such as gradient descent tend to fall into local optimum. 
It is an interesting future direction to use global search algorithms to train the deep learning models for COVID-19 diagnosis.
\end{itemize}

\begin{itemize}
\item[$\bullet$]
\textbf{Interpretability.} Deep learning has achieved excellent performance in diagnosing COVID-19. However, it is a black box, and it is difficult to understand the cause of certain predictions. Hence, explainable deep learning models are worth studying and are the key to making AI techniques truly effective.
\end{itemize}

\section{Conclusion}
Coronavirus disease 2019 (COVID-19) has significantly influenced healthcare systems and finical markets around the world since its outbreak. To interrupt the spreading virus, medical imaging has proven to be an important tool in early COVID-19 detection and severity evaluation. In this survey, we investigate the main scope and contribution of deep learning applications based on medical imaging in diagnosing COVID-19. Although deep learning model based on medical images cannot replace existing RT-PCR test at current stage, it has shown great potential in diagnosing COVID-19 and has become an important complement to RT-PCR. 
Meanwhile, we gather available datasets for diagnosing COVID-19 and point out that appropriate image preprocessing techniques can improve the generalization performance of the model. Later, we introduce the existing deep learning applications for diagnosing COVID-19, including lesion segmentation, image classification and severity quantification. Meanwhile, some methods used to improve the performance of deep model are discussed. Finally, we discuss some challenges and future directions for using deep learning technologies and medical image processing to diagnose COVID-19. We believe that with the help of deep learning and image processing technologies, and many other disciplines, the outbreak of COVID-19 will be better managed. 
We sincerely hope that this paper will be a good reference and will drive more new studies on deep learning and medical imaging to fight against COVID-19 epidemic and future outbreak on respiratory disease.


\bibliographystyle{ACM-Reference-Format}

\end{document}